\documentclass{JHEP3} 

\JHEPspecialurl{http://jhep.sissa.it/JOURNAL/JHEP3.tar.gz}

\usepackage{epsfig,multicol,bbm}

\newcommand\fverb{\setbox\fverbbox=\hbox\bgroup\verb}
\newcommand\fverbdo{\egroup\medskip\noindent%
            \fbox{\unhbox\fverbbox}\ }
\newcommand\fverbit{\egroup\item[\fbox{\unhbox\fverbbox}]}
\newbox\fverbbox

\newcommand{\lsim}{\mathrel{\mathop{\kern 0pt \rlap
  {\raise.2ex\hbox{$<$}}}
  \lower.9ex\hbox{\kern-.190em $\sim$}}}
\newcommand{\gsim}{\mathrel{\mathop{\kern 0pt \rlap
  {\raise.2ex\hbox{$>$}}}
  \lower.9ex\hbox{\kern-.190em $\sim$}}}

\newcommand{\be}{\begin{equation}}
\newcommand{\ee}{\end{equation}}
\newcommand{\beqarr}{\begin{eqnarray}}
\newcommand{\eeqarr}{\end{eqnarray}}

\title{Minimal dark matter in type III seesaw}

\author{Eung Jin Chun \\
Korea Institute for Advanced Study,
Hoegiro 87, Dongdaemun-gu, Seoul 130-722, Korea\\
Email: \email{ejchun@kias.re.kr} }

\preprint{KIAS-P09048}

\abstract{We explore the possibility of a new dark matter
candidate in the supersymmetric type III seesaw mechanism where a
neutral scalar component of the $Y$= 0 triplet can be the lightest
supersymmetric particle.  Its thermal abundance can be in the
right range if non-standard cosmology such as kination domination
is assumed.  The enhanced cross-section of the dark matter
annihilation to $W^+ W^-$ can leave detectable astrophysical and
cosmological signals whose current observational data puts a lower
bound on the dark matter mass.   The model predicts the existence
of a charged scalar  almost degenerate with the dark matter scalar
and its lifetime lies between 5.5 cm and 6.3 m. It provides a
novel opportunity of the dark mater mass measurement by
identifying slowly-moving and highly-ionizing tracks in the LHC
experiments. If the ordinary lightest supersymmetric particle is
the usual Bino, its decay leads to clean signatures of same-sign
dilepton and di-charged-scalar associated with observable
displaced vertices which are essentially background-free and can
be fully reconstructed.
 }

\begin{document}


\section{Introduction}
\label{sec:intro}

Dark matter (DM) candidates can be found in some well-motivated
extensions of Standard Model.  The best-known example would be
Minimal Supsersymmetric Standard Model (MSSM) with R-parity in
which  the lightest supersymmetric particle (LSP) turns out to be
a good DM particle whose thermal relic density is naturally  in
the right range \cite{susydm}.

Low-energy supersymmetry at the TeV scale has been motivated as a
natural explanation of the electroweak symmetry breaking, and it
may also be related to the orgin of neutrino masses. This
motivates us to introduce into MSSM a TeV-scale seesaw mechanism
explaining the observed neutrino masses and mixing. In this
regard, it is interesting to address a question whether there can
be a new dark matter candidate coming from the neutrino sector.

The most popular way to realize the seesaw mechanism
\cite{nutheory} is to introduce massive singlet (right-handed)
neutrinos (type I seesaw).  In this case, the lightest
right-handed sneutrino can be stable and become an additional
candidate of dark matter whose relic density is determined
non-thermally \cite{asaka05} or thermally if the right-handed
neutrinos carry extra $U(1)'$ interaction \cite{lee07} or Yukawa
interaction associated with the Higgs sector \cite{cerdeno08}.
Another option is to assume the presence of an $SU(2)_L$ triplet
with $U(1)_Y$ charge one ($Y=1$), which couple to two lepton
doublets and thus generate neutrino masses through its small
vacuum expectation value (type II seesaw). This model predicts
quite distinctive collider signatures through which the structure
of the neutrino mass matrix can be explored \cite{chun03}.
However, its dark matter candidate, the neutral component of the
triplet fermion, directly couples to the $Z$ gauge boson and thus
its large coupling is already ruled out by direct detection
experiments. The last option of our interest is to introduce $Y=0$
triplets which couple to lepton and Higgs doublets (type III
seesaw) \cite{type3}.  A neutral scalar component  of such triplet
superfields can be the LSP. Let us remark that such a scalar
triplet LSP can be another realization of the minimal dark matter
studied in Ref.~\cite{cirelli05}.

In the following, we explore properties of this scalar triplet
present in the supersymmetric type III seesaw mechanism as a dark
matter candidate and its signatures in collider experiments.

\section{Triplet spectrum  in supersymmetric type III
seesaw}

Type III seesaw mechanism introduces real $SU(2)_L$ triplets
$\Sigma = (\Sigma^+, \Sigma^0, \Sigma^-)$ with $Y=0$, which allows
the superpotential,
\begin{equation} \label{WIII}
 W_{III} = y_{ij} L_i H_2 \Sigma_j + {1\over2} M_k \Sigma_k
 \Sigma_k \,.
\end{equation}
Integrating out the heavy triplet fields one obtains the seesaw
neutrino mass matrix,
\begin{equation}
 {\cal M}^\nu_{ij} = y_{ik} y_{jk} {v_2^2 \over M_k}
 \,,
\end{equation}
where $v_2 = \langle H_2^0 \rangle$.  From now on, the generation
index for the triplets will be suppressed as we are interested in
the one containing the lightest neutral scalar component. The
fermion triplet components, denoted by $\Sigma^{\pm,0}$, have the
common mass $M$. For the mass spectrum of the scalar triplet
components, denoted by $\tilde{\Sigma}^{\pm, 0}$, we need to
consider the supersymmetric mass $M$, the soft supersymmetry
breaking masses and the electroweak mass splitting. The neutral
scalar components with $T_3=0$ take the masses given by
 \begin{equation} \label{mm0}
 m^2_{\tilde{\Sigma}^0_{2,1}} = M^2 +\tilde{m}^2 \pm B M
 \end{equation}
where $\tilde{m}$ is the diagonal soft mass and $B$ is the
bilinear soft term which will be assumed to be positive without
loss of generality. The charged scalar components
$\tilde{\Sigma}^\pm$ carrying $T_3=\pm1$ have the following
mass-squared matrix:
 \begin{equation} \label{matpm}
 {\cal M}^2_{\tilde{\Sigma}^\pm} = \left[
 \begin{array}{cc}
  M^2+\tilde{m}^2+c_W^2 m_Z^2 c_{2\beta} & B M \cr
  B M & M^2+\tilde{m}^2-c_W^2 m_Z^2 c_{2\beta} \cr
 \end{array} \right]\,,
 \end{equation}
where $c_W$ is the cosine of the weak mixing angle and the angle
$\beta$ is defined by $t_\beta= v_2/v_1$. Diagonalizing
Eq.~(\ref{matpm}), one finds the mass-squared eigenvalues,
 \begin{equation} \label{mmpm}
 m^2_{\tilde{\Sigma}^\pm_{2,1}} = M^2 +\tilde{m}^2 \pm
 \sqrt{B^2 M^2 + c_W^4 m_Z^4 c^2_{2\beta} }\,.
 \end{equation}
 Note that the
lighter scalars, $\tilde{\Sigma}_1^\pm$ and $\tilde{\Sigma}_1^0$,
can have smaller masses than their fermionic partners and the
ordinary superparticles ($m_{\tilde{\Sigma}_1} < M, \tilde{m}$)
due to a large negative term driven by the $B$-term, and can be
even lighter than the ordinary lightest supersymmetric particle
(OLSP) in the MSSM. The tree-level mass splitting between
$\tilde{\Sigma}_1^\pm$ and $\tilde{\Sigma}_1^0$ driven by the
electroweak symmetry breaking is given by
 \begin{equation}
 \Delta m_{\rm tree} \equiv
 [ m_{\tilde{\Sigma}_1^\pm}-m_{\tilde{\Sigma}_1^0} ]_{\rm tree}
 \approx -{c_W^2 m_Z^4 c^2_{2\beta} \over 4 B M m_{\tilde{\Sigma}_1^0}}
 \end{equation}
in the limit of $m_Z^2 \ll B M$.  Note that the tree-level mass
splitting is negative. On the other hand, one-loop correction
induces positive contribution given by
 \begin{equation}
 \Delta m_{\rm loop} = {\alpha_2 m_{\tilde{\Sigma}_1^0}\over 4\pi}
 \big[ f(r_W) - c_W^2 f(r_Z) \big]
 \end{equation}
where $r_{W,Z}=m_{W,Z}/ m_{\tilde{\Sigma}_1^0}$ and  $f(r)=-r[2
r^3 \ln r +(r^2-4)^{3/2} \ln (r^2-2-r\sqrt{r^2-4})/2]/4$
\cite{cirelli05}.

\FIGURE[t]{ \epsfig{file=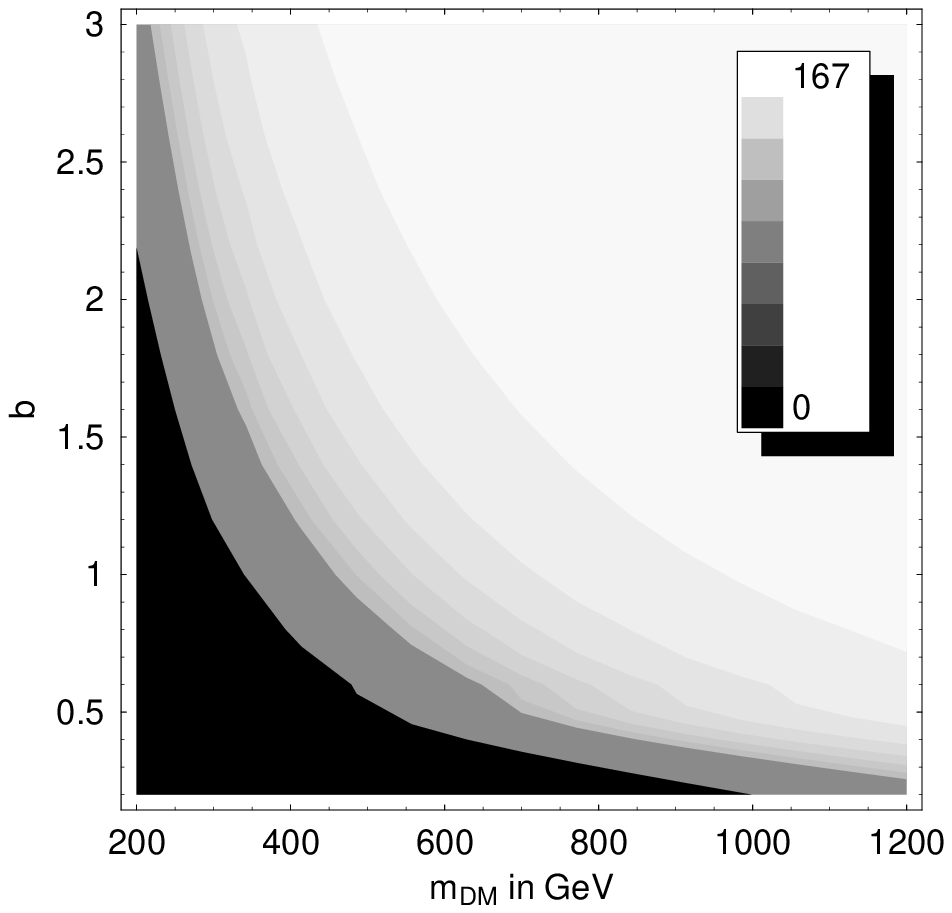,width=9cm,height=7cm}
 \caption{Contour plot of mass splitting $\Delta m$ between the charged
and the neutral (DM) scalar components as a function of  the DM
mass $m_{\rm DM}$ and the parameter $b\equiv \sqrt{BM}/m_{\rm
DM}$.  The black region is for $\Delta m < 0$ and thus excluded
for our discussion. In the next darkest region, we have $\Delta
m<100$ MeV, and $\Delta m$ is increased by 10 MeV for each contour
region. In the brightest region, $\Delta m$ is larger than 160 MeV
but below the upper limit of 167 MeV. \label{fig1}} }

The one-loop splitting reaches the typical maximum value $\Delta
m_{\rm loop} \approx 167$ MeV for $m_{\tilde{\Sigma}_1} \gg
m_{W,Z}$ which can be partly canceled by the tree-level
contribution.  In Fig.~1, we show the total mass splitting $\Delta
m = \Delta m^{\rm loop} + \Delta m^{\rm tree}$ in the plane of the
lightest neutral scalar mass $m_{\rm DM}=m_{\tilde{\Sigma}_1^0}$
and the dimensionless parameter $b\equiv \sqrt{BM}/m_{\rm DM}$
quantifying the size of the $B$-term.
 Apart from the black region in the lower left corner,
the neutral scalar component remains lighter than the charged
scalar and can be the LSP dark matter.  Depending on the size of
$\Delta m$, some interesting collider signatures will occur as
will be discussed in Section 4.

\section{Relic density  and indirect detection limits}

The dominant annihilation channel of the dark matter,
$\tilde{\Sigma}_1^0$, is the `direct' gauge coupling with
$W^{\pm}$, and thus it has a large s-wave annihilation rate
\cite{cirelli05}:
 \begin{equation} \label{Xann}
 \langle \sigma v \rangle \approx {4\pi \alpha_2^2 \over
 m^2_{\tilde{\Sigma}_1^0} } \,.
 \end{equation}
It can give rise to the right thermal relic density only in the
multi-TeV range in the standard cosmology. For smaller mass, the
annihilation cross-section becomes large to suppress the standard
thermal relic density.  In this case, the correct amount of dark matter relics
must come from a certain non-thermal origin, or from a non-standard thermal
history of the universe. An interesting possibility for the latter
is a kinetic energy (kination) domination driven by the evolution
of quintessence which forms dark energy today \cite{caldwell97}.
Such a kination era would immediately follow the period of
inflation in a cosmological scenario identifying the inflaton and
quintessence field \cite{spokoiny93}, where one can still find
sufficient reheating through gravitational particle production
\cite{ford87}.

\FIGURE[t]{ \epsfig{file=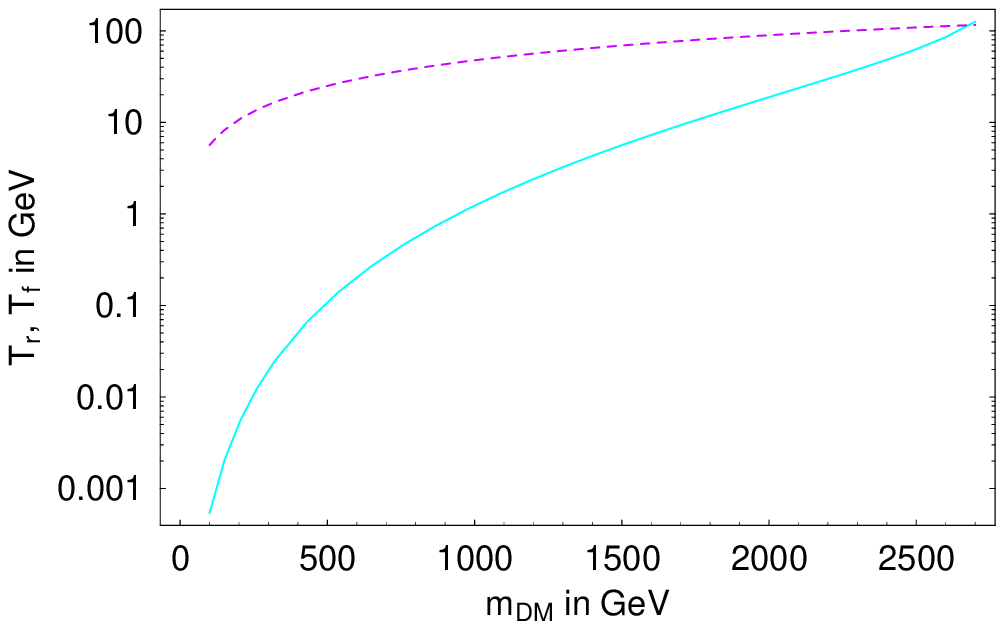,width=9cm} \caption{ The
solid line shows the reduced kination-radiation equality
temperature, $T_r\equiv \sqrt{g_{*eq}/g_{*f}}\,T_{eq}$, required
to produce the right dark matter density. The dotted line is he
freeze-out temperature $T_f$.
    \label{fig2}}
}

For a phenomenological consideration of kination cosmology, it is
enough to identify one free parameter $T_{eq}$ at which the
kination and radiation components become equal.  At a given
temperature $T$, the expansion parameter can be expressed as
 \begin{equation}
 H(T) = H_{st} \sqrt{ 1+ {g_*\over g_{*eq}}{ T^2 \over  T_{eq}^2} }
 \end{equation}
where $H_{st}$ is the Hubble parameter in the standard cosmology,
$H_{st}\approx 1.66\sqrt{g_*}\, T^2/m_{pl}$.  If dark matter
freeze-out occurs before $T_{eq}$ (when $H > H_{st}$), a larger
annihilation cross-section is required to produce a right thermal
relic density. Given the dark matter freeze-out temperature $T_f$
and the kination-radiation equality temperature $T_{eq}$, one can
find the approximate formula for the present dark matter density:
 \begin{equation}
 \Omega_{\rm DM} h^2 \approx {0.88\times10^{-10} {\rm GeV}^{-2}
 \over \langle \sigma v\rangle }{z_f\over \sqrt{g_{*f}}} k({z_r\over
 z_f})\,,
 \end{equation}
where $z_f=m_{\tilde{\Sigma}_1^0}/T_f$,
$z_r=\sqrt{g_{*f}/g_{*eq}}\, m_{\tilde{\Sigma}_1^0}/T_{eq}$, and
$k(u) = u/\ln(u+\sqrt{1+u^2})$ \cite{salati02}. Note that the
function $k$ characterizing the kination domination recovers the
standard value $k=1$ in the limit of $u\to 0$.  For the purpose of
our investigation, we calculate $T_f$ approximately by equating
the annihilation cross-section and the expansion parameter.  Then,
requiring $\Omega_{\rm DM} h^2=0.11$, one can find appropriate
values of the freeze-out temperature $T_f$ and the redefined
kination-radiation equality temperature $T_r
=\sqrt{g_{*eq}/g_{*f}}\,T_{eq}$ as functions of the dark matter
mass $m_{\tilde{\Sigma}_1^0}$.  The result is shown in Fig.~2.
 One finds that the required $T_r$ is larger than about 5 MeV for the dark
matter mass larger than 200 GeV, and thus there is no conflict
with the standard big-bang nucleosynthesis in the parameter region
of our interest.

\FIGURE[t]{ \epsfig{file=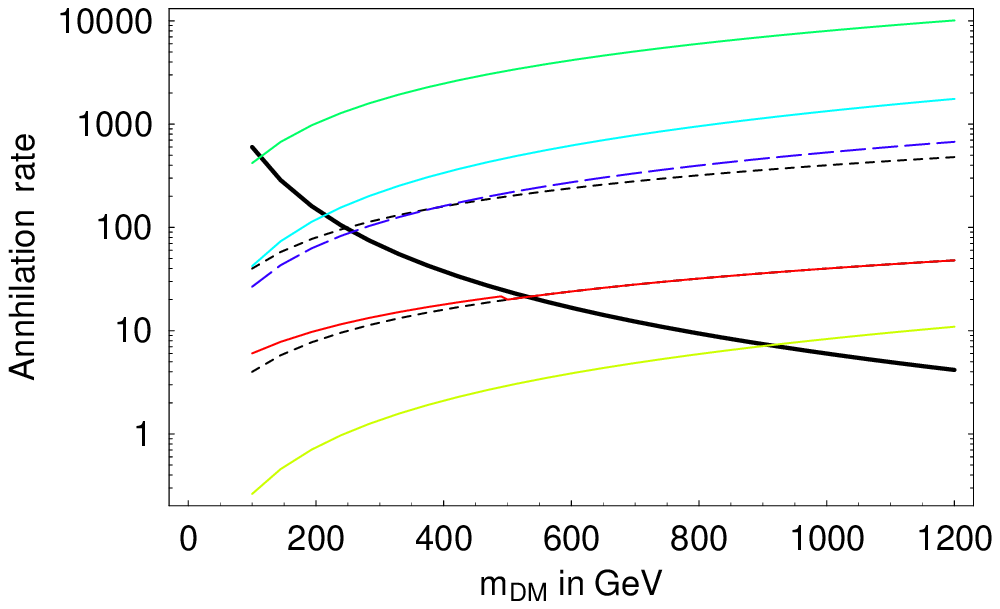,width=9cm}
\caption{ The cross-section of the scalar triplet dark matter
annihilating into $W^+ W^-$ is shown by the thick solid line
in unit of $3\times10^{-26}$ cm$^3$/sec.
Thin solid and dashed lines summarize various astrophysical and cosmological constraints.
Solid lines from the uppermost one come from the observations of extra-galactic diffuse gamma background [Belikov-Hooper \cite{ext}], big-bang nucleosynthesis
\cite{bbn}, anti-proton flux in cosmic rays \cite{pbar}, and
galactic center radio observation  (assuming NFW profile)
\cite{gc}. The upper (lower) short-dashed line is the limit from CMB
measurement by WMAP5 (PLANCK) \cite{cmb}. The long-dashed line is the optical depth
bound [Cirelli et.al. \cite{reion}].
    \label{fig3}}
}

Concerning the direct detection of our dark matter particle, let
us recall that it has $Q=Y=0$ and its leading contribution to the
scattering on nuclei comes from scalar interaction at one-loop.
Thus its nucleonic cross-section is quite small: $\sigma_{\rm
nucleon} \approx 10^{-45}$ cm$^2$
 \cite{cirelli05}, which is about two orders of magnitude below
 the current direct detection limit.  On the other hand, it is amusing to note that
our scalar dark matter triplet with $\Delta m \sim 10$ MeV can
explain the DAMA/LIBRA modulation results \cite{bai09} through the
element-dependent resonant scattering \cite{pospelov08}.

The  dark matter annihilation rate (\ref{Xann}) of
 $\tilde{\Sigma}_1^0  \tilde{\Sigma}_1^0 \to W^+ W^-$ is
much larger than the standard freeze-out value: $3\times 10^{-26}$
cm$^3$/sec for sub-TeV dark matter mass,  as shown by the thick
solid line in Fig.~3. Such an enhanced annihilation can lead to
various indirect signals in astrophysics and cosmology, and thus
gets restricted by current observational data. Fig.~3 summarizes
various constraints analyzed in the recent literature considering
the effects of dark matter annihilation to the extra-galactic
diffuse gamma-ray background by produced energetic electrons and
positrons \cite{ext}, the changes in light element abundances
predicted successfully by the big-bang nucleosynthesis \cite{bbn},
the anti-proton flux in cosmic rays \cite{pbar}, the radio
observation from the galactic center \cite{gc}, the extra
energy injection during the recombination epoch \cite{cmb}, and
the observed optical depth of the Universe \cite{reion}.
The most stringent limit comes from the galactic center radio
observation which strongly depends on astrophysics models. The NFW
profile is assumed for the lowest solid line which puts a strong
bound:
$
m_{\tilde{\Sigma}_1^0}
> 900\;\mbox{GeV}\,.
$
  But, this bound disappears if the isothermal profile
  is assumed. Therefore, we do not take this bound for our
  consideration of collider signatures.
At the moment, the most stringent bound comes from the
cosmic ray anti-proton flux measurements;
\begin{equation}
 m_{\tilde{\Sigma}_1^0} > 520 \;\mbox{ GeV},
 \end{equation}
which is comparable to the future CMB limit.
 Independently of
such indirect detection constraints, collider experiments will be
able to provide another limit on the scalar triplet mass.

\section{LHC signatures}

Some of interesting collider signatures of neutrino mass models at
TeV scale can occur in  lepton-flavor violating decays of new
particles responsible for generating light neutrino masses. They
usually involve displaced vertices related to small neutrino
masses and same-sign dilepton final states.  In type III seesaw,
the production of triplet fermions through off-shell W boson can
yield such a signal: $pp\to \Sigma^\pm \Sigma^0 \to l^\pm l^\pm
W^\mp Z^0$ \cite{hambye08,aguila08,arhrib09,li09}. The decay rate
of the triplet fermion is given by $\Gamma_\Sigma \approx y^2
M/8\pi$ where the neturino Yukawa coupling can be traded with the
effective neutrino mass: $\tilde{m}_\nu \equiv y^2 v_2^2/M$.  Thus
its lifetime becomes $\tau_\Sigma\approx 0.1$ mm for
$\tilde{m}_\nu=0.05$ eV and $M=v_2\approx 174$ GeV. This result is
applicable also to type I seesaw. In type II seesaw, where doubly
charged Higgs boson decays to two same-sign leptons, the maximum
decay length can be found to be about 0.3 mm \cite{chun03}.

In our DM scenario in supersymmetric type III seesaw, we
concentrate on the collider phenomenology of the triplet scalars.
A peculiar feature of displaced vertices reflecting small neutrino
Yukawa couplings occurs now in the decay of the ordinary lightest
supersymmetric particle. For our discussion, we will take the
usual lightest neutralno $\chi_1^0$ (mostly Bino) as the OLSP.
Note that the seesaw mechanism in the fermion (scalar) sector
induces mixing between lepton (slepton) doublets and fermion
(scalar) triplets. In particular, two-body decays of the OLSP can
arise through the mixing between sleptons and scalar triplets
induced by the soft $A$-term and supersymmetric $F$-term,
 \begin{equation}
 V_{\rm mix} =
 m_D A\,[ {1\over\sqrt{2}}\,\tilde{\nu} \tilde{\Sigma}^0
 + \tilde{l} \tilde{\Sigma}^+ ]
 + m_D M \, [{1\over\sqrt{2}}\,\tilde{\nu} \tilde{\Sigma}^{0*}
 + \tilde{l} \tilde{\Sigma}^{-*} ] + h.c. \,,
 \end{equation}
where $m_D$ denotes the Dirac neutrino mass between $\nu$ and
$\Sigma^0$: $m_D = y v_2$.  The OLSP decay rates, for $\chi_1^0
\to l^\pm \tilde{\Sigma}_1^\mp$ and $ \nu \tilde{\Sigma}_1^0$, are
found to be
\begin{eqnarray} \label{Gnlsp}
 \Gamma( \chi_1^0 \to \nu \tilde{\Sigma}_1^0) &=& {\alpha'\over
 16} \theta_{\tilde{\nu}}^2\, m_{\chi_1^0} \left(1-{m_{\tilde{\Sigma}_1
 }^2 \over m_{\chi_1^0}^2}\right)^2 ,\nonumber \\
 \Gamma( \chi_1^0 \to l^\pm \tilde{\Sigma}_1^\mp) &=& {\alpha'\over
 32} \theta_{\tilde{l}}^2\, m_{\chi_1^0} \left(1-{m_{\tilde{\Sigma}_1
 }^2 \over m_{\chi_1^0}^2}\right)^2  .
 \end{eqnarray}
Here the small parameters $\theta_{\tilde{l},\tilde{\nu}}$
quantify the mixing  between  $\tilde{l}^\pm$ ($\tilde{\nu}$) and
$\tilde{\Sigma}_1^\pm$ ($\tilde{\Sigma}_1^0$):
 \begin{equation}
 \theta_{\tilde{l},\tilde{\nu}} \approx
 {M(M-A) \over m^2_{\tilde{l},\tilde{\nu}} -
 m^2_{\tilde{\Sigma}_1}}\, \delta \quad\mbox{where}\quad
 \delta \equiv \sqrt{\tilde{m}_\nu\over M}\,.
 \end{equation}
Recall that $\delta$ is the small mixing angle between $\nu$ ($l$)
and $\Sigma^0$ ($\Sigma^-$) coming from the seesaw relation
$\tilde{m}_\nu = m_D^2/M$. Compared to the fermion triplet decay
rate $\Gamma_{\Sigma}$ discussed above, the OLSP decay rate
 (\ref{Gnlsp}), $\Gamma_{\chi^0_1} \approx \alpha' \delta^2
m_{\chi_1^0}/8$, is typically smaller by an order of magnitude,
and thus we expect to have more distinct displaced vertices.
Taking again $\theta_{\tilde{l}, \tilde{\nu}} \approx \delta$ and
$\tilde{m}_\nu = 0.05$ eV,  the OLSP decay length becomes
 \begin{equation} \label{t-olsp}
 \tau_{\chi_1^0} \approx \left[ {\alpha'\over 8}
  \tilde{m}_\nu  { m_{\chi^0_1} \over M} \right]^{-1} \approx
 0.3  \,\mbox{cm}
 \end{equation}
for $m_{\chi^0_1}=M$.

The  charged and neutral scalar fields present in our model are
nearly degenerate with each other.  As shown in Section 1, the
tree-level mass splitting driven by the $B$-term and $D$-term
masses always gives negative contribution to $\Delta m = m_{
\tilde{\Sigma}_1^\pm} - m_{ \tilde{\Sigma}_1^0}$, which  can
partly cancel the positive one-loop correction.   As a consequence
of this, the mass splitting $\Delta m$ can be smaller than the
maximal splitting of $\Delta m\approx 167$ MeV induced solely by
one-loop correction \cite{cirelli05}. Depending on whether $\Delta
m$ is larger or smaller than $m_{\pi^\pm}$, one finds the allowed
decay channel, $\tilde{\Sigma}_1^\pm \to \tilde{\Sigma}_1^0
\pi^\pm$ or $\tilde{\Sigma}_1^0 e^\pm \nu$, whose rates are given
by \cite{cirelli05}:
 \begin{eqnarray} \label{2-3-decay}
 \Gamma(\tilde{\Sigma}_1^\pm \to \tilde{\Sigma}_1^0 \pi^\pm)
  &=& {2\over \pi}\, G_F^2 V_{ud}^2 \Delta m^3 f_\pi^2
 \sqrt{1-{m_\pi^2\over \Delta m^2}} \,; \nonumber\\
 \Gamma(\tilde{\Sigma}_1^\pm \to \tilde{\Sigma}_1^0 e^\pm \nu)
  &=& {2  \over 15 \pi^3}\, G_F^2 \Delta m^5 \,.
 \end{eqnarray}
Note that we have a similar situation as in the Wino LSP scenario
\cite{winolsp} with much more restricted value of $\Delta m$.
Independently of the scalar triplet mass, the two-body decay rate
takes its maximum value for $\Delta m \approx 167$ MeV, and its
minumum value for $\Delta m = m_{\pi^\pm}$, below which the three
body decay channel opens up.  Unless the mass splitting $\Delta m$
is finely tuned to be less than $m_{e^\pm}=0.5$ MeV, the three
body decay rate of Eq.~(\ref{2-3-decay}) is applied for $\Delta m
\leq m_{\pi^\pm}$. Thus, we restrict ourselves to the decay length
of the charged scalar triplet in the range:
 \begin{equation} \label{trackrange}
   5.5 \;\mbox{cm} \lsim \;
   \Gamma^{-1}_{\tilde{\Sigma}_1^\pm} \;
   \lsim 6.3\; \mbox{m} \,.
  \end{equation}
Recall that $\pi^\pm$ or $e^\pm$ from the $\tilde{\Sigma}_1^\pm$
decay is too soft to be observed in the LHC detectors.  Thus one
can only observe highly-ionizing tracks caused by
$\tilde{\Sigma}_1^\pm$ which typically disappear somewhere inside
a detector.  When such a heavy charged particle moves slowly
($\beta<1$), its momentum and velocity (and thus mass) can be
measured simultaneously, providing a unique background-free and
model-independent search for new physics. Experimental studies on
the mass reconstruction have been performed by using the
time-of-flight measurement at the muon detectors of ATLAS and CMS,
and also by the ionization-rate (dE/dx) measurement at the tracker
of CMS \cite{atlas-cms}. Both methods can be applied to our case
if the mass splitting is close to or below $m_{\pi^\pm}$ [see
(\ref{trackrange})].

The heavy scalar triplet $\tilde{\Sigma}_1^\pm$ can be produced
following the (cascade) production of the OLSP:
 \begin{equation} \label{Xcascade}
 pp\to \chi^0_1 \chi^0_1 \to
 \cases{ l^{\pm} l^{\pm} \tilde{\Sigma}_1^\mp
 \tilde{\Sigma}_1^\mp \cr
l^{\pm} l^{\mp} \tilde{\Sigma}_1^\pm
 \tilde{\Sigma}_1^\mp \cr}\,,
 \end{equation}
or directly through $W$/$Z$ boson exchange:
 \begin{equation} \label{Xstriplet}
 p p \to \cases{ \tilde{\Sigma}_1^\pm   \tilde{\Sigma}_1^0 \cr
                 \tilde{\Sigma}_1^+  \tilde{\Sigma}_1^- \cr} \,.
 \end{equation}
In the first case, one can have both the same-sign dileptons and
same-sign anomalous charged tracks resulting from the Majorana
nature of the mother particle.  Note that the OLSP mass can be
reconstructed from the energy (momentum) measurement of the
charged leptons and the mass measurement of
$\tilde{\Sigma}_1^\pm$.  Furthermore, the reconstructed displaced
vertices of the OLSP decay (\ref{t-olsp}) can be useful to
distinguish the candidate events. The production cross-section of
the process (\ref{Xcascade}) depends on models of supersymmetry
breaking and sparticle mass spectrum.  Assuming the squark and gluino
have the same mass,  their production cross-section at the LHC
with $\sqrt{s}=14$ TeV  is 2 pb for the squark/gluino mass 1 TeV \cite{beenakker09}, providing a large number of the lepton number violating processes
and anomalous charged tracks  (\ref{Xcascade}) even at the early stage of the LHC.
From the result of Ref.~\cite{beenakker09}, we can infer that
the scalar triplet signals with a few events can be probed
up to the squark/gluino mass of 1.7 (2.8) TeV for the luminosity of 0.1 (10) fb$^{-1}$.

Apart from this,   the
process (\ref{Xstriplet}) provides another means to test our model,
independently of the squark and gluino  mass. The
corresponding cross-sections have been calculated in
Ref.~\cite{perez08}. For $m_{\tilde{\Sigma}_1}=550$ GeV, for instance,
the production cross-sections for $\tilde{\Sigma}_1^\pm
\tilde{\Sigma}_1^0$ and $\tilde{\Sigma}_1^+  \tilde{\Sigma}_1^-$
are 2.2 fb and 1.1 fb, respectively.  Thus,
copious (single and double) anomalous charged tracks can be observed if the dark matter mass is close to its lower bound and the integrated luminosity reaches a nominal value of 10 fb$^{-1}$.
The largest scalar triplet mass, for which the scenario under discussion
can be probed at the LHC luminosity of 10 (100) fb$^{-1}$, is about 800 (1250) GeV
corresponding to the total cross-section of 0.5 (0.05) fb.  We note that, given the dark matter mass bound of 520 GeV, the direct production is too small to be observed for the initial luminosity of 100 pb$^{-1}$.  However, the cascade production from the initial squark/gluino decay could be observed if the squark/gluino mass is smaller than about
1.7 TeV as mentioned above.

Let us finally make a remark on  the fermion triplet signature.
When the fermion mass $M$ is not too large, one can look for the
same-sign dilepton signal $pp \to l^\pm l^\pm + 4j$ followed by
the $\Sigma^\pm \Sigma^0$  production as mentioned earlier in this section.
Note that its production rate is 10 times larger than the scalar
production rate, and the above lepton number violating signals can be probed
up to a fermion mass 450 (700) GeV at the LHC luminosity of 10 (100) fb$^{-1}$ \cite{arhrib09}. Let us also recall that the trilepton signal
$l^\pm l^\pm l^\mp$ is as good as the same-sign dilepton signal
in probing the fermion triplet \cite{aguila08}.

\section{Conclusion}

It is suggested that the neutral scalar component
$\tilde{\Sigma}^0_1$ of the $Y$= 0 triplet in the supersymmetric
type III seesaw mechanism can be a viable dark matter candidate.
The $B$-term contribution can lead to a large mass splitting
between two neutral scalars and the lighter one becomes the LSP.
For its mass in the sub-TeV range, the annihilation cross-section
is much larger than the standard value $3\times10^{-26}$
cm$^3$/sec required for a thermal production of the observed relic
density. Such an enhanced annihilation can be consistently
accommodated by considering a non-standard cosmology: kination
domination around the dark matter freeze-out era.  This kind of
scenario could be tested or limited indirectly through various
astrophysical observations as summarized in Fig.~3. Currently the
strongest bound comes from the radio observation from the galactic
center which is however strongly dependent on the dark matter
profile. Next stringent constraint is put by the cosmic ray anti-proton
measurement, requiring $m_{\tilde{\Sigma}^0_1 } > 520$ GeV.

A peculiar aspect of our model in view of collider phenomenology
is the existence of the charged scalar particle
$\tilde{\Sigma}^\pm_1$ which is almost degenerate with the dark
matter particle $\tilde{\Sigma}^0_1$.  Their mass splitting
$\Delta m$ gets both the (negative) tree-level and (positive)
one-loop contributions which can cancel each other. When $\Delta m
< m_{\pi^\pm}$, the charged scalar allows only three-body decay,
$\tilde{\Sigma}^\pm_1 \to \tilde{\Sigma}^0_1 e^\pm \nu$, with
$\tau\approx 6.3$ m. This charged partner of the dark matter is
another example of heavy long-lived charged particles whose mass
can be measured model-independently by detecting their anomalous
tracks in the LHC experiments. The charged scalars can be produced
directly through the W/Z boson exchange, or after the production
and the decay of the ordinary lightest supersymmetric particle
like the Bino. In the latter case, we can probe clean signatures
of same-sign dilepton and di-charged-scalar tracks having
measurable displaced vertices ($\tau \sim 0.3$ cm). Furthermore,
the mass measurement of the charged scalars allows us to determine
the OLSP mass as well.

 Given the direct production
cross-section of our scalar particles, $pp \to \tilde{\Sigma}^\pm_1 \tilde{\Sigma}^0_1$
and $\tilde{\Sigma}^+_1 \tilde{\Sigma}^-_1$,
 numerous anomalous charged particle tracks
could be observed if the scalar dark matter mass is in its low end;
e.g., about 30 events
for the mass 550 GeV at the luminosity 10 fb$^{-1}$.
 The discovery/exclusion of the model can be made up
to $m_{\tilde{\Sigma}^0_1 } < 1250 $ GeV with the integrated
luminosity 100 fb$^{-1}$ as no background events are expected.
On the other hand, the previous studies showed that the direct production of the
fermion triplet $pp \to \Sigma^+ \Sigma^0$ leading to the same-sign dilepton
signals can be probed up to the fermion triplet mass 700 GeV.
 Of
course,  one could observe much more events if a squark or gluino
and thus the OLSP can be copiously produced, which depends on the
mass spectrum of heavier supersymmetric particles.

\medskip

{\bf Acknowlegement:} The author thanks Suyong Choi and Manuel
Drees for valuable comments on the LHC signals.  This work was
supported by Korea Neutrino Research Center through National
Research Foundation of Korea Grant (2009-0083526).



\begin{thebibliography}{99}

\bibitem{susydm}
 For a review, see, G. Jungman, M. Kamionkowski and K. Griest,
  Phys.\ Rept.\  {\bf 267}, 195 (1996)
  [arXiv:hep-ph/9506380].

\bibitem{nutheory}
 For a review, see, R.~N.~Mohapatra {\it et al.},
  ``Theory of neutrinos: A white paper,''
  Rept.\ Prog.\ Phys.\  {\bf 70}, 1757 (2007)
  [arXiv:hep-ph/0510213].


\bibitem{asaka05}
 T. Asaka, K. Ishiwata and T. Moroi,
  Phys.\ Rev.\  D {\bf 73}, 051301 (2006)
  [arXiv:hep-ph/0512118].
\bibitem{lee07}
 H.-S. Lee, K. T. Matchev and S. Nasri,
  Phys.\ Rev.\  D {\bf 76}, 041302 (2007)
  [arXiv:hep-ph/0702223].
\bibitem{cerdeno08}
 D. G. Cerdeno, C. Munoz and O. Seto,
  Phys.\ Rev.\  D {\bf 79}, 023510 (2009)
  [arXiv:0807.3029 [hep-ph]];
  F.~Deppisch and A.~Pilaftsis,
  JHEP {\bf 0810}, 080 (2008)
  [arXiv:0808.0490 [hep-ph]].

\bibitem{chun03}
  E.~J.~Chun, K.~Y.~Lee and S.~C.~Park,
  Phys.\ Lett.\  B {\bf 566}, 142 (2003)
  [arXiv:hep-ph/0304069].
\bibitem{type3}
  R. Foot, H. Lew, X.-G. He and G. C. Joshi, Z.~Phys.~ C44 (1989)
  441.
\bibitem{cirelli05}
 M.~Cirelli, N.~Fornengo and A.~Strumia,
  Nucl.\ Phys.\  B {\bf 753}, 178 (2006)
  [arXiv:hep-ph/0512090].

\bibitem{caldwell97}
  R. R. Caldwell, R. Dave  and P. J. Steinhardt,
  {\it Phys.\ Rev.\ Lett.}\  {\bf 80} 1582 (1998)
  [arXiv:astro-ph/9708069];
  P. J. Steinhardt, L. M. Wang and  I. Zlatev,
  {\it Phys.\ Rev.\  D} {\bf 59} 123504 (1999)
  [arXiv:astro-ph/9812313].

\bibitem{spokoiny93}
 B. Spokoiny,
 {\it  Phys.\ Lett.\  B} {\bf 315} 40
  [arXiv:gr-qc/9306008];
 P. J. E. Peebles  and A. Vilenkin,
 {\it  Phys.\ Rev.\  D} {\bf 59} 063505 (1999)
  [arXiv:astro-ph/9810509].

\bibitem{ford87}
 L.~H.~Ford,
  Phys.\ Rev.\  D {\bf 35}, 2955 (1987);
   E.~J.~Chun, S.~Scopel and I.~Zaballa,
  JCAP {\bf 0907}, 022 (2009)
  [arXiv:0904.0675 [hep-ph]].



\bibitem{salati02}
  P.~Salati,
  Phys.\ Lett.\  B {\bf 571}, 121 (2003)
  [arXiv:astro-ph/0207396].

\bibitem{bai09}
 Y. Bai and P. J. Fox, arXiv:0909.2900 [hep-ph].
\bibitem{pospelov08}
  M.~Pospelov and A.~Ritz,
  Phys.\ Rev.\  D {\bf 78}, 055003 (2008)
  [arXiv:0803.2251 [hep-ph]].


\bibitem{ext}
 M.~Kawasaki, K.~Kohri and K.~Nakayama,
  Phys.\ Rev.\  D {\bf 80}, 023517 (2009)
  [arXiv:0904.3626 [astro-ph.CO]];
  S.~Profumo and T.~E.~Jeltema,
  JCAP {\bf 0907}, 020 (2009)
  [arXiv:0906.0001 [astro-ph.CO]];
  A.~V.~Belikov and D.~Hooper,
  arXiv:0906.2251 [astro-ph.CO].
\bibitem{bbn}
 J.~Hisano, M.~Kawasaki, K.~Kohri, T.~Moroi and K.~Nakayama,
  Phys.\ Rev.\  D {\bf 79}, 083522 (2009)
  [arXiv:0901.3582 [hep-ph]].
\bibitem{pbar}
  F.~Donato, D.~Maurin, P.~Brun, T.~Delahaye and P.~Salati,
  Phys.\ Rev.\ Lett.\  {\bf 102}, 071301 (2009)
  [arXiv:0810.5292 [astro-ph]].
\bibitem{gc}
 G.~Bertone, M.~Cirelli, A.~Strumia and M.~Taoso,
  JCAP {\bf 0903}, 009 (2009)
  [arXiv:0811.3744 [astro-ph]].
\bibitem{cmb}
 S.~Galli, F.~Iocco, G.~Bertone and A.~Melchiorri,
  Phys.\ Rev.\  D {\bf 80}, 023505 (2009)
  [arXiv:0905.0003 [astro-ph.CO]];
  T.~R.~Slatyer, N.~Padmanabhan and D.~P.~Finkbeiner,
  arXiv:0906.1197 [astro-ph.CO].
\bibitem{reion}
 G. Huetsi, A. Hektor and M. Raidal, arXiv:0906.4550 [astro-ph.CO];
 M. Cirelli, F. Iocco and P. Panci, arXiv:0907.0719 [astro-ph.CO].


\bibitem{hambye08}
 R.~Franceschini, T.~Hambye and A.~Strumia,
  Phys.\ Rev.\  D {\bf 78}, 033002 (2008)
  [arXiv:0805.1613 [hep-ph]].
\bibitem{aguila08}
  F.~del Aguila and J.~A.~Aguilar-Saavedra,
  Nucl.\ Phys.\  B {\bf 813}, 22 (2009)
  [arXiv:0808.2468 [hep-ph]].
\bibitem{arhrib09}
  A.~Arhrib, B.~Bajc, D.~K.~Ghosh, T.~Han, G.~Y.~Huang, I.~Puljak and G.~Senjanovic,
  arXiv:0904.2390 [hep-ph].
\bibitem{li09}
  T.~Li and X.~G.~He,
  arXiv:0907.4193 [hep-ph].


\bibitem{winolsp}
 C.~H.~Chen, M.~Drees and J.~F.~Gunion,
  Phys.\ Rev.\ Lett.\  {\bf 76}, 2002 (1996)
  [arXiv:hep-ph/9512230];
J.~L.~Feng, T.~Moroi, L.~Randall, M.~Strassler and S.~f.~Su,
  Phys.\ Rev.\ Lett.\  {\bf 83}, 1731 (1999)
  [arXiv:hep-ph/9904250].


\bibitem{perez08}
 P.~Fileviez Perez, H.~H.~Patel, M.~J.~Ramsey-Musolf and K.~Wang,
  Phys.\ Rev.\  D {\bf 79}, 055024 (2009)
  [arXiv:0811.3957 [hep-ph]].

\bibitem{atlas-cms}
S. Giagu, ``Search for long lived particles in ATLAS and CMS'',
 Talk presented at ICHEP 2008, Philadelphia, PA;
The ATLAS Collaboration, ``Expected Performance of the ATLAS
 Experiment : Detector, Trigger and Physics'', CERN-OPEN-2008-020
 [arXiv:0901.0512 [hep-ex]];
The CMS Collaboration,
 ``Search for heavy stable charged particles
 with 100/pb and 1/fb in the CMS experiment'', CMS-PAS-EXO-08-003.


\bibitem{beenakker09}
  W.~Beenakker, S.~Brensing, M.~Kramer, A.~Kulesza, E.~Laenen and I.~Niessen,
  arXiv:0909.4418 [hep-ph].


\end{thebibliography}
\end{document}